\documentclass[reprint,prl,aps,floatfix,unsortedaddress]{revtex4-1}
\usepackage{graphicx}
\usepackage{amsmath,amsfonts,mathcomp}
\usepackage[modulo]{lineno}
\usepackage{siunitx}
\usepackage{xcolor,colortbl,tabularx,multirow}
\usepackage{listings}
\usepackage[caption=false]{subfig}
\usepackage{array}
\newcolumntype{L}[1]{>{\raggedright\let\newline\\\arraybackslash\hspace{0pt}}m{#1}}
\newcolumntype{C}[1]{>{\centering\let\newline\\\arraybackslash\hspace{0pt}}m{#1}}
\newcolumntype{R}[1]{>{\raggedleft\let\newline\\\arraybackslash\hspace{0pt}}m{#1}}

\begin{document}
\title{Computational solvation dynamics of \textcolor{black}{oxyquinolinium betaine} linked to trehalose}
\author{Esther Heid and Christian Schr\"oder}
\email{christian.schroeder@univie.ac.at}
\affiliation{University of Vienna, Faculty of Chemistry, Department of Computational Biological Chemistry, 
W\"ahringerstra{\ss}e 19, A-1090 Vienna, Austria}

\begin{abstract}
Studying the changed water dynamics in the hydration layers of biomolecules is an important step towards fuller understanding of their function and mechanisms, but has shown to be quite difficult. 
The measurement of the time-dependent Stokes shift of a chromophore attached to the biomolecule is a promising method to achieve this goal, as published 
in \textit{J.~Phys.~Chem.~Lett.}, \textbf{5} (2014), 1845  where trehalose was used as biomolecule, 1-methyl-6-\textcolor{black}{oxyquinolinium} betaine as chromophore and water as solvent. A overall retardation
of solvent molecules is then obtained by comparison of the linked system to the same system without trehalose, but contributions from different subgroups
of solvent molecules, for example molecules close to or far from trehalose, are unknown. The difficulty arising from these unknown contributions of retarded and possibly unretarded solvent molecules is
overcome in this work by conducting computer simulations on this system and decomposing the overall signal into contributions from various molecules at different locations. We performed 
non-equilibrium molecular dynamics simulation  using a polarizable water model and a non-polarizable solute model
and could reproduce the experimental time-dependent Stokes shift accurately for the linked trehalose-\textcolor{black}{oxyquinolinium} and the pure \textcolor{black}{oxyquinolinium} over a wide temperature range,
indicating the correctness
of our employed models. Decomposition of the shift into contributions from different solvent subgroups showed that the amplitude of the measured shift is made up only half by the desired 
retarded solvent molecules in the hydration layer, but to another half
by unretarded bulk water, so that measured relaxation times of the overall Stokes shift are only a lower boundary for the true relaxation times in the hydration layer of trehalose.
As a side effect, the results on the effect of trehalose on solvation dynamics contributes to the long standing debate on the range of influence of trehalose on water dynamics, 
the number of retarded solvent molecules and the observed
retardation factor when compared to bulk water.
\end{abstract}
\maketitle

\section{Introduction}
The disaccharide trehalose ($\alpha$-D-glucopyranosyl-$\alpha$-D-glucopyranoside) is known to protect cells in some microorganisms and plants 
from injury upon desiccation~\cite{cro75a,beh97a,col93b}, freezing~\cite{som96a,sto90a} or heating~\cite{tim97a,hay98a,mey98a,bha03a} and has therefore attracted lots of interest during the last decades. It was shown to protect proteins and 
membranes and to preserve the content and structure of liposomes~\cite{cor97a,hav08a}. 
Different mechanisms have been \textcolor{black}{suggested, although none of them could be proved:} 
The early model of water replacement around proteins~\cite{hal12a} and biomembranes~\cite{cro71a} was contradicted by recent studies, which 
\textcolor{black}{suggested} that trehalose is mainly excluded from protein surfaces, so that a general 
description of the biofunction of trehalose via this approach has become rather improbable~\cite{swe16a} and may only occur close to membranes~\cite{sar16a}.
Another theory states that the exceptionally high glass-transition temperature of trehalose and the high viscosity of concentrated trehalose solutions \textcolor{black}{may}
suppress crystallization~\cite{sci97a} and even induce glass transitions in the cytoplasm~\cite{hal12a}. 
Evidence was also found that trehalose influences the solvation dynamics of the biomolecular hydration layer and consequently the protein dynamics~\cite{hav08a,swe16a,gal13a}. 
Heyden \textit{et al.} reported that the effectiveness of bioprotection of different carbohydrates increases with increasing ability of slowing down water dynamics
 and concluded 
that solvent dynamics retardation is a key aspect of desiccation protectants~\cite{hav08a}. 
\textcolor{black}{The bioprotection of trehalose is therefore thought to be connected to the induced changes in static and dynamic behavior of hydration shell water, which is consistent
with the second and third theory mentioned above.
Yet, a complete description of the physical mechanism of structure protection is still missing~\cite{oth16a}. }
Furthermore, there is still disagreement on the range and 
nature of changes in the solvent network induced by the presence of trehalose~\cite{hal12a}.
\textcolor{black}{This work therefore contributes to this debate by
simulating trehalose in water, yielding the range and temperature dependence of solute-induced changes of solvent properties. 
To this aim, the chromophore 1-methyl-6-oxyquinolinium betaine (1MQ) connected to trehalose, as described by Ernsting and coworkers in Ref.~\cite{ern14a}, 
serves as a probe for the local solvation dynamics around trehalose via the time-dependent Stokes shift.}

\bigskip
\textcolor{black}{The time-dependent Stokes shift describes the transient behavior of solvent relaxation after an electric perturbation realized via an electronic excitation of a dissolved chromophore 
and is therefore a local measure of solvent properties close to the chromophore. This locality and the ability of measuring dynamic and not only static solvent properties makes time-dependent fluorescence
spectroscopy a promising method to complement measurements of solvation shell properties via neutron diffraction, conventional terahertz spectroscopy or NMR spectroscopy.
After optical excitation of the chromophore, the emitted fluorescence spectrum
shifts from higher to lower wavenumbers as the solvent molecules rearrange to stabilize the excited state and its new electronic distribution}. 
The normalized time evolution of the corresponding \textcolor{black}{wavenumber} $\nu(t)$
\begin{equation}
 S(t)=\frac{\nu(t)-\nu(\infty)}{\nu(0)-\nu(\infty)}
\end{equation}
is called the \textcolor{black}{relaxation function $S(t)$ of the time-dependent Stokes shift}  and is (mainly) a measure of solvent properties. Measurements and simulations of $S(t)$ in different solvents were initially
applied to investigate the nature and timescales of bulk solvent relaxation, which are an important factor to chemical reaction rates in solution~\cite{fle87a,cas95a,mar93a,kri97a}. 
\textcolor{black}{The timescale of relaxation of the time-dependent Stokes shift was also used to characterize solvent properties in hydration shells, for example close to proteins~\cite{hal05a,qih15a,sen15a,sen16a}, 
DNA~\cite{ber09a} or carbohydrates~\cite{ern10a,ern14a,oth16a} as it enables the local measurement of hydration shell properties without measuring almost exclusively bulk water. However, only recent 
development of fluorescence spectroscopy made it possible to measure $\nu(t)$ accurately enough
to characterize the small changes in $S(t)$ inflicted by some solute changes or upon temperature change, which opens up the exciting possibility of detecting even small changes in solvent properties induced by biomolecules,
and comparing results quantitatively to THz spectra as seen in Ref.~\cite{ern14a} .}

\bigskip
Albeit these promising studies, there are also some drawbacks connected with the interpretation of the experimental time-dependent Stokes shift, as it resembles only the sum of multiple processes that happen after
excitation of the solute. For example, molecules at different
spatial locations contribute to the overall shift, but the contributions cannot be disentangled experimentally. 
In the 1MQ close to trehalose system of Ernsting and coworkers \cite{ern14a}, 
the solute probes the solvent molecules in its vicinity, where some of the solvent molecules might be close to the disaccharide and therefore show changed properties in comparison to 
bulk water, but some also might be far from the disaccharide and not affected by its presence.
The ratio of these contributing solvent molecules is, to the best of our knowledge, unknown, but is an indispensable information when interpreting experimental data, 
as large unexpected contributions to the overall shift from bulk water might tamper the accuracy of experimental results and conclusion drawn from them.
Furthermore, the probe molecule only scans 
a small area, so that part of the induced changes by the trehalose moiety in water dynamics cannot be resolved.
This could be problematic, as
\textcolor{black}{closer investigation} of the trehalose-water system showed that
subpopulations in the sugar hydration shell exhibit a significantly different dynamic behavior, both translational and rotational, so that the influence of trehalose on water dynamics is 
site-dependent and therefore heterogeneous~\cite{cam11a,hal12a}.
 It is therefore necessary to elucidate the ongoing molecular processes, which is possible via computer simulation
of the system to validate and interpret experimental data.
\textcolor{black}{In this sense, this work studies the dynamical changes in water induced by trehalose according to Ref.~\cite{ern14a} more thoroughly via computer simulation.
The time-dependent Stokes shift relaxation function of the 
fluorescent probe 1MQ is decomposed 
into contributions from 
water molecules at different distances to the trehalose and oxyquinolinium moiety,
which enables the calculation of the range of trehalose-induced changes in water dynamics. Additionally, 
 the effect of temperature change on the Stokes shift and
 the effect of the linkage of trehalose to the chromophore on the observed Stokes shift is investigated. 
 The extent of solvent retardation at different temperatures will provide information on whether the same mechanisms apply for protection of proteins at high and low temperatures via trehalose.}
Furthermore, the contribution of trehalose motion to the overall shift is calculated.
Thus, this work contributes to a better understanding of the induced changes in water dynamics by trehalose and assesses the significance of measuring  hydration shell properties  via the time-dependent Stokes shift.

\section{Methods}
In all our simulations reported here, the water solvent is represented by the polarizable SWM4 water model~\cite{rou03a} 
which reproduces quite well experimental dynamical behavior. In order to have a reliable description of the solvation of the chromophore 
all simulation boxes of \textcolor{black}{trehalose linked to oxyquinolinium (referred to as 1TQ)} contain 2000 water molecules. Simulations of 1MQ contain 1000 water molecules.
This way, the majority of the water molecules will behave bulk like. 
The force field and in particular the partial charge distribution of 1MQ is described in detail in Ref.~\cite{sch16b}
and therefore only briefly summarized here. The partial charge distribution of 1MQ in ground and excited state was taken from quantum-chemical calculations using TD-DFT with 
the $\omega$B97xD functional~\cite{cha08a} and the aug-cc-pVTZ basis set in a polarizable continuum model (PCM)~\cite{tom05b}.
The trehalose attached to the \textcolor{black}{oxyquinolinium} betaine is modeled using a fully atomistic, non-polarizable force field obtained from 
PARAMCHEM~\cite{van12a,van12b} and the CHARMM General Force Field~\cite{van10a}. The partial charges for the \textcolor{black}{oxyquinolinium} part of 1TQ are the same as 1MQ 
whereas partial charges for the sugar moiety were taken from PARAMCHEM. Neutral net charge of the combined trehalose-\textcolor{black}{oxyquinolinium system} was achieved by modifying the charge of the linking CH$_2$-unit.
Furthermore, we assumed the partial charges of the attached trehalose not to change upon solute excitation as the excitation involves only the $\pi$-system of the \textcolor{black}{oxyquinolinium}.
\begin{figure}[ht]
 \centering
 \subfloat[][extended configuration]{\includegraphics [width=0.52\linewidth]{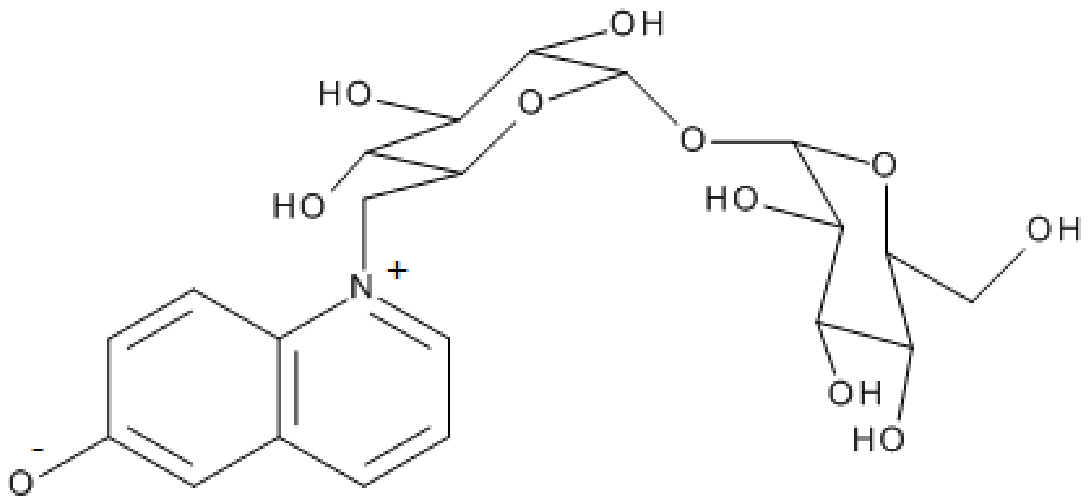}}
  \subfloat[][folded configuration]{\includegraphics [width=0.40\linewidth]{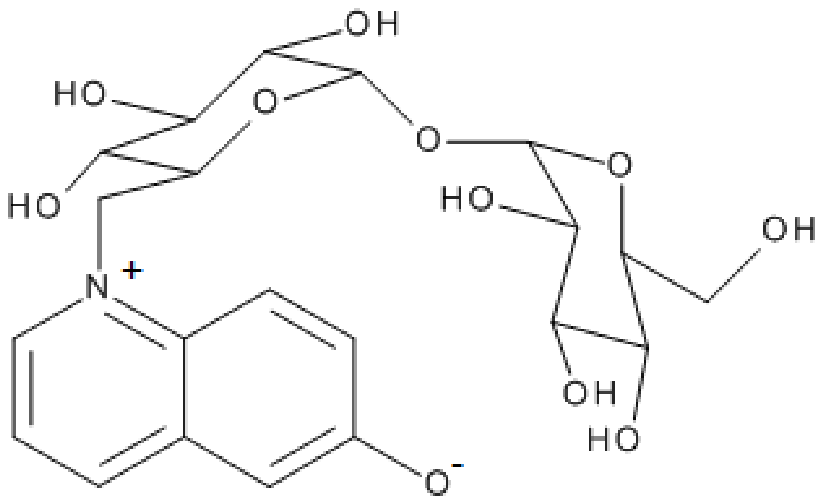}}
  \caption{Principal configurations of trehalose linked to \textcolor{black}{oxyquinolinium betaine}}
  \label{FIG:1TQ}
\end{figure}
Since the rotation around the linking CH$_2$-unit of 1TQ is probably slower than the observation window of the computational Stokes shifts, we performed two
independent series of 500 non-equilibrium MD simulations starting at the extended and folded configuration at \SI{20}{\celsius} as depicted in Fig.~\ref{FIG:1TQ}, respectively.
These series are accompanied by additional 500 non-equilibrium simulations of 1MQ in an aqueous solution of \SI{0.4}{M} trehalose (16 molecules trehalose and 2000 molecules SWM4). 
The extended 1TQ and 1MQ were simulated at temperatures of \SI{4}{\celsius}, \SI{20}{\celsius}, \SI{27}{\celsius} and \SI{60}{\celsius}. As the trajectories for both 1TQ and 1MQ at \SI{20}{\celsius} were
used to dissect the overall Stokes shift into molecular contributions, the number of trajectories for these two systems was increased to 1000 to improve statistics.
Each system was randomly packed in a cubic box and equilibrated in CHARMM~\cite{kar09a} using a NpT-ensemble at the respective temperature and a 
pressure of \SI{1}{bar} until the box length converged to the values shown in Table~\ref{TAB:boxl}. 
\begin{table}
 \centering
 \caption{Converged boxlengths of the simulated systems}
 \begin{tabular}{lC{1cm}C{1cm}C{1cm}C{1cm}}
 \hline\hline\
  &\multicolumn{4}{c}{Boxlength [\AA]}\\
 &\SI{4}{\celsius}&\SI{20}{\celsius}&\SI{27}{\celsius}&\SI{60}{\celsius}\\ 
\hline
 1MQ &31.04&31.14&31.20&31.48\\
 1TQ &39.10&39.25&39.33&39.67\\
 1MQ+Treh. &&40.38\\
   \hline\hline
 \end{tabular}
 \label{TAB:boxl} 
\end{table}

\bigskip
The independent starting configurations were obtained from a NVT simulation at elevated temperature.
Each configuration was equilibrated during a \SI{0.5}{\nano\second} NVT run, then excited by abruptly changing the partial charge distribution and afterwards monitored for \SI{50}{\pico\second}.
For all simulations periodic boundary conditions were used, with the Particle Mesh Ewald method to calculate interaction energies (grid size \SI{1}{\angstrom}, cubic splines of order 6,
$\kappa$ of \SI{0.41}{\angstrom}$^{-1}$, energy cut-off at \SI{11}{\angstrom}).
The resulting trajectories were used to calculate the average interaction energies $\Delta \bar U$, from which the time-dependent Stokes shift relaxation functions can be calculated as
\begin{equation}
 S(t)=\frac{\Delta \bar U(t)-\Delta \bar U(\infty)}{\Delta \bar U(0)-\Delta \bar U(\infty)} 
\end{equation}
where $\Delta \bar U$ is the change in Coulomb interaction energy upon excitation averaged over all trajectories
 \begin{equation}
\Delta U(t) = \frac{1}{4 \pi \epsilon_0} \sum\limits_{\gamma} \sum\limits_{i\beta} \frac{\Delta q_{\gamma} \cdot q_{i\beta}}{r_{\gamma i\beta}(t)}
\label{EQU:DeltaU}
\end{equation}
between the chromophore atoms $\gamma$ of the solute molecule 
and the solvent atoms $\beta$ of molecule $i$ at distance $r_{\gamma i\beta}$ where  $\Delta q_{j\gamma}$ corresponds to the change of the partial charge distribution
 from ground $\mathbb{S}_0$ to excited state $\mathbb{S}_1$. 
Furthermore we used parameter-free Voronoi analysis~\cite{ste09d,oka00a} to assign the solvent molecules to different shells around the solute molecule. We then decomposed the overall
relaxation function to contributions of the solvent shells. A Python program based on MDAnalysis~\cite{bec11a} performed the calculation of Voronoi shells and Stokes shifts.
The resulting shifts were fitted via sums of Gaussian and stretched exponential functions
\begin{equation}
 S(t) \approx a e^{-(t/\tau_1)^2} + (1-a) e^{-(t/\tau_2)^\beta}
 \label{EQU:fit}
\end{equation}
yielding average relaxation times
\begin{equation}
 \langle\tau\rangle = a \frac{\tau_1}{2} \sqrt{\pi} + (1-a) \frac{\tau_2}{\beta} \Gamma \Bigl(\frac{1}{\beta} \Bigl).
 \label{EQU:tau}
\end{equation}
where $\Gamma$ means the gamma-function. Also, a distance based analysis was conducted to dissect the overall \textcolor{black}{relaxation function} into contributions from water molecules at different distances to the surface of the 
disaccharide to explore the range of retardation stemming from the trehalose moiety. Mean residence times in the first shell around 1MQ and 1TQ, as well as a hydrogen bond analysis around trehalose,
were additionally computed to account for the type of interaction between trehalose and water. 
The criteria for a hydrogen bond was chosen to be a maximum distance of \SI{2.4}{\angstrom} between oxygen and hydrogen, and an angle of 
more than \SI{120}{\degree} between oxygen (acceptor), hydrogen and oxygen (donor) as reported in Ref~\cite{des11a}.

\section{Results and discussion}
\subsection{Effect of the linkage of trehalose to 1MQ}
To investigate whether the covalent bond between trehalose
and \textcolor{black}{oxyquinolinium} influences the solvation dynamics of the surrounding solvent molecules and to check the general validity of our system,
we calculated the  Stokes shift at \SI{20}{\celsius} of 1MQ in water and 0.4M trehalose aqueous solution, 
as well as 1TQ (extended configuration) in water as
shown in Fig.~\ref{FIG:neq_exp}. For reasons of clarity, the actual data points in this -- and also in the next figure -- 
were omitted and only the fitting functions plotted. Actual data points, as well as the full data range can be seen later in Fig.~\ref{FIG:neq_contr} or Fig.~\ref{FIG:neq_gau}.
\begin{figure}[t]
 \centering
 \includegraphics [width=\linewidth]{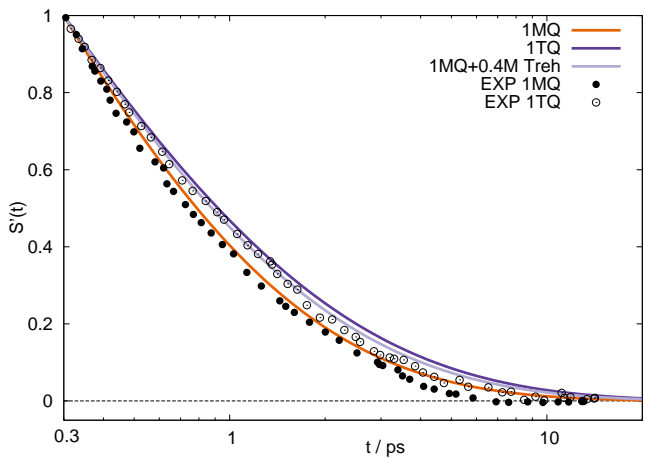}
 \caption{Fitting functions of the scaled \textcolor{black}{time-dependent Stokes shift relaxation functions} of 1MQ (orange) and 1TQ (dark violet) in SWM4 water compared to experimental data extracted from Ref.~\cite{ern14a}, 
 as well as 1MQ in a 0.4M trehalose solution (light violet).}
 \label{FIG:neq_exp}
\end{figure}
\textcolor{black}{The corresponding measurements from Ernsting and coworkers~\cite{ern14a} of 1MQ and 1TQ are also shown.}
As the experimental \textcolor{black}{relaxation curve} was set to 1 at \SI{0.3}{ps} our data was rescaled
accordingly, so that 
\begin{equation}
 S'(t)=\frac{\Delta \bar U(t)-\Delta \bar U(\infty)}{\Delta \bar U(0.3)-\Delta \bar U(\infty)} = \frac{S(t)}{S(0.3)}.
\end{equation}
The MD simulation reproduces the experimental curve very well, indicating that the applied model for 1TQ yields reasonable results. The validity of the pure \textcolor{black}{oxyquinolinium} and SWM4 model was already shown in Ref.~\cite{sch16b}.
When comparing the relaxation dynamics of 1MQ to 1TQ, it can be seen that the trehalose moiety slows down the solvation dynamics of water. 
For 1MQ, a relaxation time of \SI{0.28}{ps} was calculated, and is raised to \SI{0.42}{ps} upon inclusion of
the disaccharide.
When comparing 1TQ to 1MQ in 0.4M trehalose solution (16.2~wt\%), nearly identical overall relaxation behavior of the Stokes shift is observed. 
This is in good agreement with corresponding measurements from Ernsting and coworkers~\cite{ern14a} who also found the Stokes shift \textcolor{black}{relaxation function} of 1TQ in water nearly indistinguishable from 1MQ in 0.4M
aqueous trehalose solution.

 We also calculated the Stokes shift for the folded conformation 1TQ at \SI{20}{\celsius} and compared it to the above used model of 1TQ but could not find any differences (not shown).  
Measurement/simulation of the relaxation behavior of the hydration layer of trehalose or other biomolecules can therefore be achieved in two different ways, as long as only the overall Stokes shift is of interest:
On the one hand, measurements can be conducted at very high concentration of the biomolecule, so that nearly no bulk water is present and the measured signal only corresponds to the hydration layer. This
approach was applied in some former spectroscopic measurements~\cite{hav08a,hal12a,des11a,fio12a} and corresponds to our simulation of 1MQ in the quite concentrated 16.2~wt\% trehalose solution. On the other hand, measurements at natural, 
low concentrations become possible
through the linkage of the chromophore to the biomolecules of interest, corresponding to our 1TQ simulation in pure water. 
As both simulations, 1TQ in water and 1MQ in 0.4M trehalose solution, show a similar Stokes shift, we can conclude that the linkage between trehalose and the chromophore itself does not influence the 
shift, but only ensures that 1MQ comes close enough to trehalose to actually probe the hydration layer of trehalose instead of bulk water.

However, on the molecular level there should be differences in the processes leading to the overall relaxation behavior between the folded 1TQ, the extended 1TQ and the 1MQ/0.4M trehalose system. 
A detailed analysis on these differences will be given later. For now, after
having confirmed that our 1TQ model gives correct predictions regarding the Stokes shift when compared to experimental data, we can  go on to explore 
the contribution of trehalose motion to the overall shift, as well as
temperature effects.

\subsection{Contribution of trehalose motion to the \textcolor{black}{Stokes shift relaxation function}}
\begin{figure}[t]
 \centering
 \includegraphics [width=\linewidth]{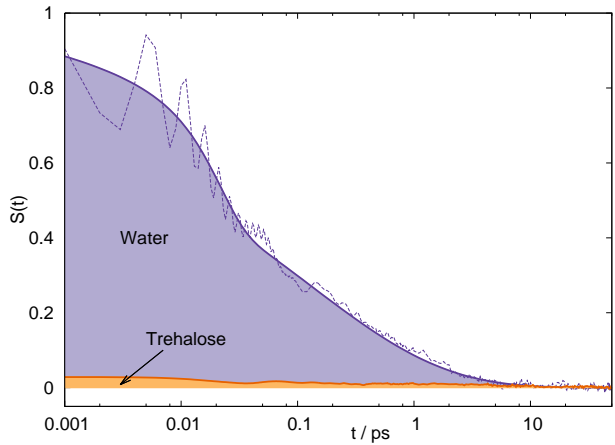}
 \caption{Contribution of trehalose movement (orange) to the overall Stokes shift \textcolor{black}{relaxation function} (violet) of 1TQ in SWM4 water at \SI{20}{\celsius}.}
 \label{FIG:neq_contr}
\end{figure}

So far, movement of the trehalose moiety was not taken into account when calculating the time series of solvation energies. To check the contribution of trehalose motion
to the overall shift, we included trehalose in the energy calculation (see Eq.~\eqref{EQU:DeltaU}) as 'solvent molecule' and then decomposed the relaxation into water and trehalose contributions. The movement of  trehalose
only marginally influences the magnitude of the Stokes shift. Without inclusion of the sugar to the energy calculation, the overall Stokes shift is \SI{4350}{cm}$^{-1}$,
and rises slightly upon inclusion to \SI{4490}{cm}$^{-1}$, which is nearly the same as in the 1MQ-SWM4 system.
However, the relaxation time does not change, and after respective normalization of the relaxation curve, no difference 
in behavior could be observed. Fig.~\ref{FIG:neq_contr} shows graphically the contribution of trehalose motion to the energy calculation.
As the trehalose contribution to the relaxation behavior of the Stokes shift is negligible, the slower relaxation dynamics of 1TQ compared to 1MQ indeed stems only from the water molecules
that are slowed down via long and short range interactions of the disaccharide.

\subsection{Temperature dependence of the solvation dynamics}
\begin{figure}[t]
 \centering
 \includegraphics [width=\linewidth]{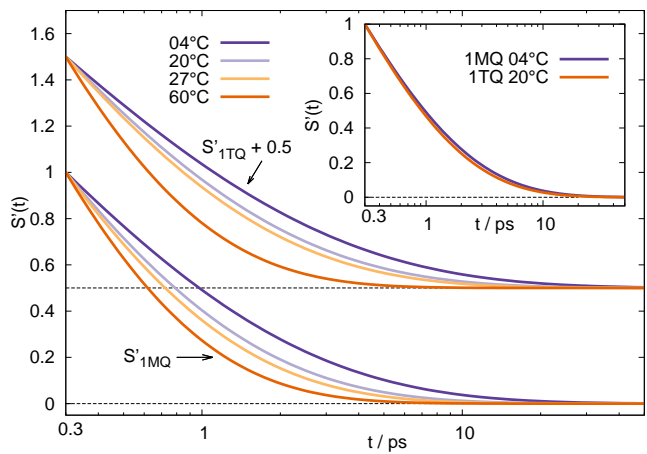}
 \caption{Fitting functions of the scaled time-dependent Stokes shift \textcolor{black}{relaxation functions} of 1MQ  and 1TQ (shifted by 0.5) in SWM4 water at \SI{4}{\celsius} (dark violet), 
 \SI{20}{\celsius} (light violet), \SI{27}{\celsius} (light orange) and \SI{60}{\celsius} (dark orange). Inset: Comparison of 1MQ at \SI{4}{\celsius}  and 1TQ at \SI{20}{\celsius}.}
 \label{FIG:neq_temp}
\end{figure}

The group of Ernsting reported that at \SI{4}{\celsius} 1MQ behaves similar to 1TQ at \SI{22}{\celsius}~\cite{ern14a}, indicating that the slowing down of the water dynamics through trehalose is similar
to a drastic cooling down of water.  We therefore decided 
to investigate the effect of temperature on the Stokes shift of 1MQ and 1TQ and 
conducted additional simulations at \SI{4}{\celsius}, \SI{27}{\celsius} and \SI{60}{\celsius}. As \textcolor{black}{$S(t)$} depends on the viscosity of the solvent we calculated the
diffusion coefficients of SWM4 water at the respective temperatures and compared them to experiment~\cite{sac00a} as shown in Table~\ref{TAB-diff} and found reasonable agreement.
The corresponding Stokes shift \textcolor{black}{relaxation functions} of 1MQ and 1TQ at the four different temperatures are depicted in Fig~\ref{FIG:neq_temp}.
\begin{table}[b]
 \centering
 \caption{Comparison of calculated (sim.) and measured (exp.) diffusion coefficients of water.}
 \begin{tabular}{llrrrr}
 \hline\hline
 $T$ [\SI{}{\celsius}]			&&4&20& 27&60\\\hline
 $D$ [$10^{-9}$ $\frac{\mathrm{m}^2}{\mathrm{s}}$]	&sim.& 1.44 &2.25&2.59& 4.77\\
					&exp.& 1.27&2.03& 2.30&4.75\\
   \hline\hline
 \end{tabular}
 \label{TAB-diff} 
\end{table}
Like already found by experiment, the relaxation time of 1MQ at \SI{4}{\celsius} is approximately the 
relaxation time of 1TQ at \SI{20}{\celsius} as visible in Table~\ref{TAB-relax} and the inset in Fig.~\ref{FIG:neq_temp}. The slowing down of water molecules in the vicinity of trehalose 
is therefore comparable to cooling down of the solvent, at least in respect to the overall relaxation behavior of the Stokes shift.

It should be noted that the relaxation times $\langle \tau \rangle$ of 1MQ scale roughly with the diffusion coefficient of water at the respective temperatures. 
At low temperatures, the difference in relaxation times between 1MQ and 1TQ is largest, whereas at high temperatures the difference vanishes completely. The retardation of water molecules
through the trehalose moiety can thus be overcome if the temperature (the kinetic energy) is large enough. 
\textcolor{black}{The cryoprotectant ability of trehalose therefore seems to be connected with the large slowing down of water dynamics at low temperatures. However, trehalose is also known to protect proteins 
at elevated temperatures, where we could find no retardation of solvation dynamics.}
Table~\ref{TAB-relax} also lists relaxation times of 1MQ and 1TQ measured in D$_2$O from Ref.~\cite{ern14a}. For \SI{60}{\celsius} they also find that 1MQ and 1TQ behave similar. At \SI{4}{\celsius}, however,
they find a drastic slowing down of relaxation dynamics for 1TQ, which  could not be detected in our light water SWM4 model. We therefore conclude that the effect is due to the deuteration of water.

\begin{figure}[t]
 \centering
 \includegraphics [width=\linewidth]{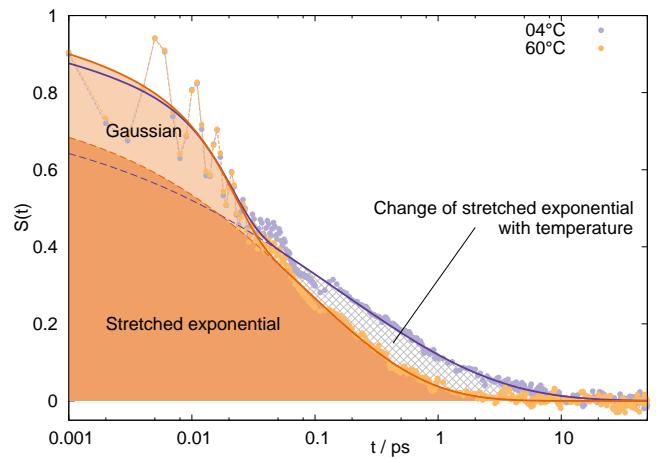}
 \caption{Dissection of the fitting function into Gaussian and stretched exponential part for 1TQ at \SI{4}{\celsius} and \SI{60}{\celsius}. Note that upon temperature change only the stretched exponential varies greatly.
 The gaussian function does not vary in shape or height.}
 \label{FIG:neq_gau}
\end{figure}

The parameters of fitting according to Eq.~\eqref{EQU:fit} for all temperatures are also listed in Table~\ref{TAB-relax}. One interesting feature is that upon temperature change, the initial Gaussian ($\tau_1$) stays the same,
whereas the stretched exponential ($\tau_2$ and $\beta$) is slowed down or fastened up as depicted in Fig. \ref{FIG:neq_gau} for 1TQ at \SI{4}{\celsius} and \SI{60}{\celsius}. The same effect is observed for 1MQ (not shown).
\begin{table}[ht!]
 \centering
 \caption{\textcolor{black}{Relaxation times using Eq.~\eqref{EQU:tau} and fitting parameters of 1MQ and 1TQ via fitting to Eq.~\eqref{EQU:fit}, 
 as well as  experimental (exp) data measured in H$_2$O and D$_2$O  from Ref.~\cite{ern14a}. 
 $\langle\tau\rangle_\mathrm{exp}$ was calculated by fitting to a single KWW function $e^{-(x/\tau_\mathrm{exp})^{\beta_\mathrm{exp}}}$, 
 and using the relation $\langle\tau\rangle_\mathrm{exp}=\frac{\tau_\mathrm{exp}}{\beta_\mathrm{exp}}\Gamma\bigl(\frac{1}{\beta_\mathrm{exp}}\bigl)$ }}
 \begin{tabular}{rR{0.2cm}cccccR{0.2cm}cc}
  \hline\hline
			&&a	&$\tau_1$ [ps]	&$\tau_2$ [ps]	&$\beta$&$\langle\tau\rangle$ [ps]&& \multicolumn{2}{c}{$\langle\tau\rangle_\mathrm{exp}$} [ps] \\\hline \underline{1MQ:}&&&&&&&&H$_2$O&D$_2$O\\[0.2cm]
  \SI{4}{\celsius}	&&0.25&0.021&0.13&0.35&\textbf{0.48}&&0.67&1.09\\
  20/\SI{22}{\celsius}	&&0.24&0.023&0.10&0.39&\textbf{0.28}&&0.43&0.42\\
  \SI{27}{\celsius}	&&0.22&0.021&0.10&0.41&\textbf{0.24}&&-&-\\
  \SI{60}{\celsius}	&&0.23&0.021&0.09&0.46&\textbf{0.16}&&-&0.38\\\hline \underline{1TQ:}&&&&&&&&H$_2$O&D$_2$O\\[0.2cm]
  \SI{4}{\celsius}	&&0.23&0.021&0.16&0.34&\textbf{0.69}&&-&2.31\\
  20/\SI{22}{\celsius}	&&0.26&0.021&0.13&0.37&\textbf{0.42}&&0.57&1.04\\
  \SI{27}{\celsius}	&&0.23&0.022&0.12&0.38&\textbf{0.35}&&-&-\\
  \SI{60}{\celsius}	&&0.22&0.022&0.08&0.45&\textbf{0.17}&&-&0.33\\  
  \hline\hline
 \end{tabular}
  \label{TAB-relax} 
\end{table}
It was shown in several studies that $S(t)$ consists of a fast Gaussian shaped part which is usually attributed to inertial motion, and therefore independent of temperature or solute, 
and an exponential/stretched exponential part attributed to diffusive rotational and translational motion~\cite{jan10a,mar93a,mar94a,mar12b,pet04a, sam02a, sam03a}. Our observation
that for both, 1MQ and 1TQ, only the stretched exponential part changes upon temperature change therefore confirms this finding.

\subsection{Contributions of solvation shells to the overall Stokes shift \textcolor{black}{relaxation function}}
\begin{figure}[t]
 \centering
 \includegraphics [width=\linewidth]{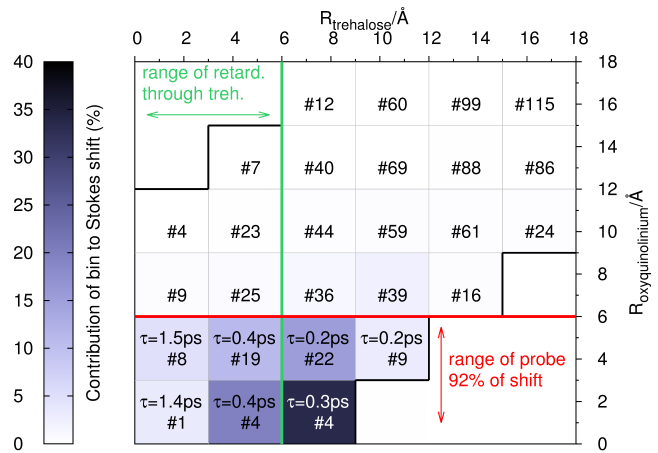}
 \caption{Contribution to the Stokes shift in \% of water molecules at distance $R_{\mathrm{trehalose}}$ to the closest trehalose atom and \textcolor{black}{$R_{\mathrm{oxyquinolinium}}$} to the closest \textcolor{black}{oxyquinolinium} atom.
 The diagram also shows the average number of solvent molecules contributing to the respective bin (indicated by '\# number') and a estimate of the respective
 relaxation time $\tau$. Bins containing no number are not occupied, which is the case for all bins outside the black thick lines.}
 \label{FIG:neq_perc}
\end{figure}

To explore the origin of the slower relaxation behavior of water close to trehalose more thoroughly, individual contributions of solvation shells around trehalose and the 
\textcolor{black}{oxyquinolinium} part were calculated using Voronoi tessellation, as well as a distance based analysis, on the \SI{20}{\celsius} 1TQ data and 1MQ data. For 1MQ 86.9\% of the Stokes shift comes from the first solvation shell
around the chromophore, 10.2\% from the second, 2.6\% from the third and 0.3\% from all other shells. To calculate this data, the interaction energy of each molecule at every time step was counted into the respective
bin based on the current shell. 
We may therefore conclude that the \textcolor{black}{oxyquinolinium} de facto probes  solvent molecules
that are either in the first or second shell, in this system about 170 water molecules altogether. 
This accounts to an averaged distance of \SI{7}{\angstrom} from the surface of the \textcolor{black}{oxyquinolinium}, and only solvent molecules closer than that radius can be probed.
As a matter of course, introducing the trehalose moiety reduces the number of solvent molecules 
that are close to the \textcolor{black}{oxyquinolinium} part of the solute, and therefore less solvent molecules contribute to the overall shift. The absolute static Stokes shift decreases from about
\SI{4450}{cm}$^{-1}$ for 1MQ to \SI{4350}{cm}$^{-1}$ for 1TQ. To estimate the thickness of the layer of retarded solvent molecules around trehalose, we decomposed the Stokes shift \textcolor{black}{relaxation function} into contributions
from water molecules at different distances to the surface of the disaccharide and the \textcolor{black}{oxyquinolinium} moiety.
\textcolor{black}{These distances are defined as the distances of the center of mass of the solvent molecule to the closest trehalose or oxyquinolinium atom, respectively.}
The analysis is purely distance based and does not use Voronoi shells. Analogously to the former analysis, the interaction
energy of a water molecule was added to the respective histogram bin based on the distance to trehalose and the chromophore for each time step separately. 
Different histogram spacings all led to the same picture, 
namely that the Stokes shift is primarily made up of water molecules at a distance of up to \SI{12}{\angstrom} to the trehalose moiety
as shown exemplarily
in Fig.~\ref{FIG:neq_perc} for a bin width of \textcolor{black}{\SI{3}{\angstrom}}.
Please keep in mind that this is the distance to the trehalose moiety not to the \textcolor{black}{oxyquinolinium} part which is shorter as we have already seen.
The x-axis resembles the distance of a water molecule to the trehalose moiety, the y-axis the distance to the \textcolor{black}{oxyquinolinium} moiety. From the plot, the former mentioned distance of probing around \textcolor{black}{oxyquinolinium}
can be read: All histogram bins under the red line sum up to about 92\% of the Stokes shift. All other water molecules contribute only marginally to the overall height of the shift.
We also calculated the relaxation times of the water molecules in the respective histogram bins and found that
 water molecules close to the sugar are heavily slowed down in their dynamics, \textit{e.g.} the relaxation time $\tau$ increases to \textcolor{black}{\SI{1.5}{ps} for the 8 water molecules which are at the surface of trehalose and
 between 3 and \SI{6}{\angstrom} apart from the oxyquinolinium moiety}. The relaxation times were calculated via Eq.~\eqref{EQU:tau} after fitting according to Eq.~\eqref{EQU:fit}, where
 only relaxation times of bins with a contribution of at least 4\% to the overall shift showed good enough statistics for the fitting procedure.
 We found that about \textcolor{black}{22} water molecules  very close to trehalose (less than \textcolor{black}{\SI{3}{\angstrom}}, first column of  Fig.~\ref{FIG:neq_perc}), are slowed down by a factor of about 4-5 compared to the 
 1MQ in water. Because of their vicinity to the trehalose surface and their strong retardation in respect to
 relaxation after an electronic perturbation we expect \textcolor{black}{some of them} to be hydrogen-bonded to the sugar. A detailed analysis of residence times and hydrogen bonds will be given later.
 About \textcolor{black}{80} molecules can be found at distances ranging from \textcolor{black}{\SI{3}{\angstrom} to \SI{6}{\angstrom}} from the trehalose moiety (second column of  Fig.~\ref{FIG:neq_perc}),
 from which about \textcolor{black}{24} molecules are close enough to oxyquinolinium to be examined. They show a 
retardation factor of about \textcolor{black}{1.4} compared to 1MQ.  All these retarded solvent molecules (first and second column) contribute only about half of the Stokes shift. Molecules further apart from the sugar part 
of 1TQ, but close enough to the 
 oxyquinolinium part to be measured, show a similar relaxation time than in the pure 1MQ-SWM4 system and are therefore not affected by the presence of trehalose. 
 These unretarded water molecules contribute the missing half of the Stokes shift.
 The Stokes shift of 1TQ consists therefore of two different signals that contribute nearly equally to the total shift: A slow component describing the heavily and and slightly retarded solvent at 
 distances below \textcolor{black}{\SI{6}{\angstrom}} from sugar, 
 and a fast component at larger distances where the solvent behaves bulk like. Such contributions cannot be disentangled from measurements alone, so that the obtained experimental Stokes shift
 is always a combination of both. A measurement of solvation dynamics purely made up of hydration shell molecules of trehalose is therefore not possible via the time-dependent Stokes shift, and the obtained
 relaxation time is only a lower boundary of the true relaxation time of hydration shell water around the sugar moiety.

 From Fig.~\ref{FIG:neq_perc} the range of retardation induced by trehalose could be calculated to extend to about 
\textcolor{black}{\SI{6}{\angstrom}}, measured from the closest trehalose atom 
 to the center of mass of the respective solvent molecules, which corresponds to about two solvation layers. Our findings are in good agreement to experimental results:
 Ernsting and coworkers found that the thickness of the hydration layer of trehalose is about \SI{5}{\angstrom} for \SI{0.4}{M} trehalose
solution, and  larger in dilute solutions~\cite{ern14a}
via measurements of the time-dependent Stokes shift at different trehalose concentrations. Heyden \textit{et al.} measured the hydration layer of trehalose via terahertz absorption measurements
(which is directly connected to the time-dependent
Stokes shift via dielectric continuum theory)
to extend 6-\SI{7}{\angstrom} from the surface of the carbohydrate ~\cite{hav08a}. Gallo and coworkers found via computer simulation and calculation of the Fourier transform of a density-density self correlation function
that trehalose influences solvation dynamics up to the second solvation shell. It should be noted, that the number of influenced solvent molecules and therefore the range of perturbation around trehalose
depends on the experimental setup and that results from static time-averaged methods, dynamic methods and computer simulation are therefore only partly comparable~\cite{fio12a,bak15a,oga15a}. Especially 
static methods underestimate the number of retarded solvent molecules, as only the strongly retarded molecules at the surface of the sugar are taken into account and the weakly retarded molecules
in the second solvation layer are neglected~\cite{oga15a}. As retardation factors are usually given as an average over all hydrated water molecules, the resulting retardation factor differs considerably with the number of retarded
solvent molecules. Halle and coworkers for example found a retardation factor of 1.6 for 47 water molecules via NMR measurements~\cite{hal12a}, whereas Lupi \textit{et al.} found a retardation factor of about 5 to 6 for only 25 
water molecules~\cite{fio12a}.
Our distance based analysis of retardation factors, which is largely independent from the calculated number of hydration water molecules, is here especially helpful, as it verifies both the high retardation factors close to trehalose,
as well as smaller retardation factors when averaging also over solvent molecules further apart from the disaccharide.

\subsection{Analysis of mean residence times and hydrogen bonds}
\begin{figure}[t]
 \centering
 \includegraphics [width=\linewidth]{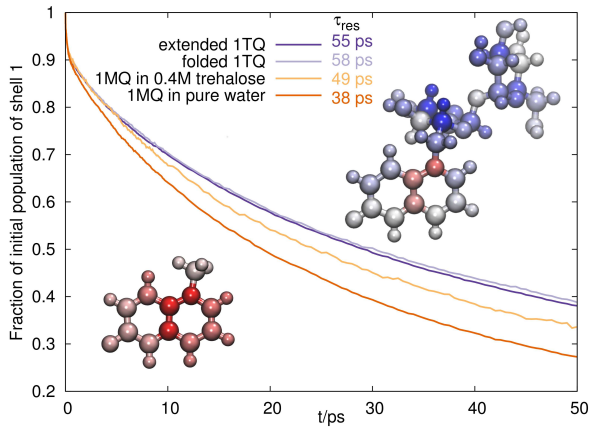}
 \caption{Fraction of initial first shell water molecules that are in shell 1 at time t for 1TQ (extended and folded conformation) and 1MQ (in 0.4M trehalose solution and in pure water).
 The mean residence times $\langle\tau_\mathrm{res}\rangle$ at the top of the figure were obtained by multi-exponential fitting. Atomistic residence times are shown in the insets, at the bottom left for 1MQ in water,
 and at the top right for 1TQ in water, where red means short residence times and blue long residence times.}
 \label{FIG:neq_res}
\end{figure}

Further insight into the change in water properties induced by trehalose is gained by inspection of first-shell residence times.
Fig.\ref{FIG:neq_res} shows the probability of finding a molecules that was initially in the first Voronoi shell of 1TQ, extended 1TQ, 1MQ in water or 1MQ in trehalose solution still (or again) in the first shell. 
From multi-exponential fitting, mean residence times $\langle\tau_\mathrm{res}\rangle$ could be
calculated  (which should not be confused with the mean relaxation time $\langle\tau\rangle$). 
It is evident, that the inclusion of the trehalose moiety slows down the exchange process around the solute as the residence time rises from about \SI{38}{ps} for the 1MQ-water system to about 55-\SI{58}{ps} for the extended 
and folded 1TQ-water system. Also, the inclusion of free trehalose in the 1MQ-0.4M trehalose system
slows down the first shell solvent exchange around 1MQ to about \SI{49}{ps}.
So far, only the residence times in respect to the whole surface of the solute was taken into account, which makes a direct comparison of the residence times around only the \textcolor{black}{oxyquinolinium} part impossible. 
However, only the slowing down of residence times around the chromophore can be linked to the slowing down in relaxation times of the Stokes shift, as solvent molecules far from the \textcolor{black}{oxyquinolinium} contribute only marginally to the shift. To get further insight into the involved atomic sites, we therefore recalculated the analysis in a 
semi-atomistic approach. The chromophore molecule was decomposed into small sections containing one non-hydrogen atom each 
plus all the attached hydrogen atoms, apart from the oxygen atom in \textcolor{black}{oxyquinolinium}, which was put together with the respective ring-carbon atom in a section. 
The probability of finding a solvent molecule that was initially in the first shell of a section still (or again) in the first shell was calculated for each section separately. 
This way, 1MQ could be described by 11 sections and 1TQ by 31 sections, so that by comparison of the \textcolor{black}{oxyquinolinium} section general conclusions about the influence of trehalose on solvent properties can be drawn.
The result is shown in the inset in Fig.\ref{FIG:neq_res}, where a red colored atom means a short residence time at that site, and blue a long residence time.
The residence times, both atomistic and overall, for 1MQ are quite short, which means that first shell solvent molecules can exchange  fast. 
The absolute residence times are of no importance, as they depend on the volume of the respective section, so that they become meaningful 
only upon comparison to other \textcolor{black}{oxyquinolinium} moieties. When comparing 1TQ to 1MQ, all chromophore sites show a slower first shell exchange, where those 
close to trehalose are slowed down more (by a factor of about 1.7) than those far from trehalose (factor of about 1.3). Averaged over all sites, a retardation 
of residence times of 1.5 is observed. The chromophore therefore shows a heterogeneous slowing-down of first-shell residence times in the presence of trehalose. 
The same scheme is observed for the folded configuration of 1TQ (not shown), where the only difference is that the trehalose is close to different \textcolor{black}{oxyquinolinium}
sites compared to the extended configuration of 1TQ (closer to the phenyl part of the chromophore). The averaged residence time is slowed down again by a factor
of about 1.5, and again, the retardation is very heterogeneous.
A different picture arises for the atomistic analysis of residence times in the 1MQ-0.4M trehalose system. All the \textcolor{black}{oxyquinolinium} sites are homogeneously 
slowed down by a factor of about 1.3, indicating a random distribution of trehalose molecules around the chromophore. The retardation factor of 1.3 
corresponds to the slightly slowing down of the \textcolor{black}{oxyquinolinium} sites in 1TQ at large distances (about two solvation shells) from the sugar. We may therefore 
conclude, that in the 1MQ-0.4M trehalose system the trehalose molecules do not come as close to the chromophore as in 1TQ. As the Stokes shift \textcolor{black}{relaxation function} in both 
systems is the same, more than one trehalose molecules has to be in vicinity to \textcolor{black}{oxyquinolinium} to make up for the weaker interaction. By measuring the Stokes 
shift of 1TQ in water and 1MQ in trehalose solution, completely different schemes of retardation are probed: Regarding the former, the chromophore probe is 
heterogeneously influenced by the presence of trehalose, and the probed solvent molecules show a spectrum of different retardation factors (both in residence 
and relaxation times). In the latter system, in contrast, the chromophore is subjected to a weaker, but homogeneous influence, so that the observed residence 
and relaxation times of solvent molecules at different
directions from the chromophore are (averaged over all simulations) equal.  

To check whether the longer overall (not atomistic) first shell residence times and slower Stokes shift relaxation of water molecules in the vicinity of trehalose could be attributed to hydrogen bonding, we 
calculated the hydrogen bonding ability of trehalose and found that all seven hydroxyl groups act as strong hydrogen-bond acceptors. Also, the oxygen ring atom  sometimes acts as 
a hydrogen-bond acceptor. The ability of trehalose to act as hydrogen-bond donor, instead of acceptor, is rather low, and the possibility of finding a hydroxyl hydrogen involved in a hydrogen bond is less than 10\%.
Overall, this leads to an average of about 9 molecules that are directly interacting with trehalose via hydrogen bonds, which is in good agreement with hydrogen bond analysis of free trehalose as reported in~\cite{cam11a}.
As it was shown before in Fig.~\ref{FIG:neq_contr}, about 
a hundred solvent molecules show slower solvation dynamics
in the presence of trehalose (compared to the solvation dynamics around 1MQ), about \textcolor{black}{twenty} of them even very slow dynamics. The large retardation of the closest molecules may therefore 
be attributed in parts to hydrogen bonding, but the observed general slowing down of water dynamics cannot be explained by the few present hydrogen bonds.
Trehalose therefore slows down water dynamics beyond those molecules directly hydrogen-bonded to it and induced a long range change in the solvation dynamics of water.

\section{Conclusion}
In this work we set up an atomistic model for trehalose covalently linked to 1-methyl-6-\textcolor{black}{oxyquinolinium} betaine to simulate solvation dynamics in water, which showed to reproduce the experimental Stokes shift well. 
We found that the characteristic slowing down of water molecules in the presence of trehalose observed by the time-dependent Stokes shift is not influenced by the linkage between the chromophore and the disaccharide and that a system of the chromophore in 0.4M trehalose solution showed similar properties regarding the overall shift.
Furthermore, different configurations of the trehalose-\textcolor{black}{oxyquinolinium} adduct led to the same relaxation behavior. 
Inclusion of trehalose movement into the relaxation process after excitation showed that the contribution is negligible. The measured Stokes shift therefore corresponds mostly to the motion of water
molecules and reflects accurately solvent behavior in this system.
In all simulated systems, the presence of trehalose, free or linked, led to retarded solvent dynamics. 
In fact, water in proximity to trehalose at \SI{20}{\celsius} behaved similar to pure water at
\SI{4}{\celsius}, as was also shown experimentally in Ref.~\cite{ern14a}, indicating a large slowing down of diffusive motion. 
The ultrafast first solvent response, which is usually assigned to inertial motion, was neither influenced
by the presence of trehalose, nor by a change in temperature. The slower response, attributed to different kinds of diffusive motion, changed upon the presence of trehalose or temperature changes. 
The inclusion of trehalose to the system therefore changes the diffusive properties of the water molecules around it, similar to what a temperature change would do.
At high temperatures (here \SI{60}{\celsius}),
the presence of trehalose did not influence the solvation dynamics of water anymore. This means that the retardation of water through trehalose can be overcome by kinetic effects. 
\textcolor{black}{The ability of trehalose to protect proteins from unfolding at high temperatures, where we could not find any solvent retardation, therefore
seems to follow a different mechanism than for its cryoprotectant ability, where solvation dynamics is retarded at low temperatures.}

Decomposition of the overall Stokes shift relaxation function into its molecular contributions gave further insights into the various processes taking place after excitation. \textcolor{black}{Oxyquinolinium} probes 
about two solvation shells, containing water molecules at different distances to trehalose, so that 
the probed solvent molecules around 1TQ can be separated into two classes: Fast unretarded molecules far apart from the trehalose moiety make up about half of the amplitude of 
the shift and are not influenced by the presence of trehalose. These
solvent molecules show similar relaxation dynamics as around 1MQ. The second class consists of very slow molecules close to trehalose which make up the second half of the amplitude of the shift. 
The molecules closest to the disaccharide (closer than \textcolor{black}{\SI{3}{\angstrom}})
are heavily slowed down, by a factor of about 4-5 compared to solvent relaxation around 1MQ, whereas solvent molecules at a distance of \textcolor{black}{\SI{3}{\angstrom} to \SI{6}{\angstrom} are only slowed down by a factor of about 1.4}. 
It should therefore be kept in mind, that experimental Stokes shifts of such system always reflect only the average over these two classes (heavily and slightly retarded 
(retardation factors \textcolor{black}{1.4} to 5) and unretarded molecules), and that the actual relaxation of molecules affected by the sugar moiety is much 
slower than the mean relaxation time of the overall measured shift. The measurement of the Stokes shift in this system therefore does  not completely reflect the desired 
dynamics (namely those in the hydration shell of trehalose) and results drawn from such measurements should be interpreted accordingly. 

We furthermore found that \textcolor{black}{about half of the} water molecules that are heavily retarded are also hydrogen bonded and concluded that the strong retardation stems from those hydrogen bonds.
The weaker retardation of the molecules farther apart, in contrast, stems from changes in the solvent network induced by trehalose possibly via the changed properties of the water molecules at the sugar surface. We also found that
hydrogen bonds are preferably build with the trehalose hydroxyl oxygen acting as acceptor (and not the hydroxyl hydrogen as a donor) and only seldomly with the three ether oxygens. 
Diverse findings on the hydrogen bonding sites in trehalose have been published
during the last five years. Pagnotta \textit{et al.} used neutron diffraction measurements with isotopic substitution and a Monte Carlo simulation to derive that most of the hydrogen bonds are at the saccharide ring
oxygens and the glycosidic oxygen~\cite{ric10a}, whereas Verde \textit{et al.} found in a computer simulation that the ether oxygens in trehalose do not form hydrogen bonds, and water molecules close to them rather from
hydrogen bonds with the closest hydroxyl oxygens, which are more electronegative~\cite{cam11a}. With our hydrogen bond analysis, the latter finding is reinforced.

\section{Acknowledgement}
This work was funded by the Austrian Science Fund FWF in the context of Project 
No. FWF-P28556-N34.



\begin{thebibliography}{53}
\expandafter\ifx\csname natexlab\endcsname\relax\def\natexlab#1{#1}\fi
\expandafter\ifx\csname bibnamefont\endcsname\relax
  \def\bibnamefont#1{#1}\fi
\expandafter\ifx\csname bibfnamefont\endcsname\relax
  \def\bibfnamefont#1{#1}\fi
\expandafter\ifx\csname citenamefont\endcsname\relax
  \def\citenamefont#1{#1}\fi
\expandafter\ifx\csname url\endcsname\relax
  \def\url#1{\texttt{#1}}\fi
\expandafter\ifx\csname urlprefix\endcsname\relax\def\urlprefix{URL }\fi
\providecommand{\bibinfo}[2]{#2}
\providecommand{\eprint}[2][]{\url{#2}}

\bibitem[{\citenamefont{Madin and Crowe}(1975)}]{cro75a}
\bibinfo{author}{\bibfnamefont{K.~A.~C.} \bibnamefont{Madin}} \bibnamefont{and}
  \bibinfo{author}{\bibfnamefont{J.~H.} \bibnamefont{Crowe}},
  \bibinfo{journal}{J. Exp. Zool} \textbf{\bibinfo{volume}{193}},
  \bibinfo{pages}{335} (\bibinfo{year}{1975}).

\bibitem[{\citenamefont{Behm}(1997)}]{beh97a}
\bibinfo{author}{\bibfnamefont{C.~A.} \bibnamefont{Behm}},
  \bibinfo{journal}{Int. J. Parasitol.} \textbf{\bibinfo{volume}{27}},
  \bibinfo{pages}{215} (\bibinfo{year}{1997}).

\bibitem[{\citenamefont{Newman et~al.}(1993)\citenamefont{Newman, Ring, and
  Colaco}}]{col93b}
\bibinfo{author}{\bibfnamefont{Y.~M.} \bibnamefont{Newman}},
  \bibinfo{author}{\bibfnamefont{S.~G.} \bibnamefont{Ring}}, \bibnamefont{and}
  \bibinfo{author}{\bibfnamefont{C.}~\bibnamefont{Colaco}},
  \bibinfo{journal}{Biotechnol. Genet. Eng. Rev.}
  \textbf{\bibinfo{volume}{11}}, \bibinfo{pages}{263} (\bibinfo{year}{1993}).

\bibitem[{\citenamefont{S{\o}mme}(1996)}]{som96a}
\bibinfo{author}{\bibfnamefont{L.}~\bibnamefont{S{\o}mme}},
  \bibinfo{journal}{Eur. J. Entomol.} \textbf{\bibinfo{volume}{93}},
  \bibinfo{pages}{349} (\bibinfo{year}{1996}).

\bibitem[{\citenamefont{Storey and Storey}(1990)}]{sto90a}
\bibinfo{author}{\bibfnamefont{K.~B.} \bibnamefont{Storey}} \bibnamefont{and}
  \bibinfo{author}{\bibfnamefont{J.~M.} \bibnamefont{Storey}},
  \bibinfo{journal}{Sci. Am.} \textbf{\bibinfo{volume}{263}},
  \bibinfo{pages}{92} (\bibinfo{year}{1990}).

\bibitem[{\citenamefont{Xie and {Timasheff}}(1997)}]{tim97a}
\bibinfo{author}{\bibfnamefont{G.}~\bibnamefont{Xie}} \bibnamefont{and}
  \bibinfo{author}{\bibfnamefont{S.~N.} \bibnamefont{{Timasheff}}},
  \bibinfo{journal}{Biophys. Chem.} \textbf{\bibinfo{volume}{64}},
  \bibinfo{pages}{25} (\bibinfo{year}{1997}).

\bibitem[{\citenamefont{Carninci et~al.}(1998)\citenamefont{Carninci,
  Nishiyama, Westover, Itoh, Nagaoka, Sasaki, Okazaki, Muramatsu, and
  Hayashizaki}}]{hay98a}
\bibinfo{author}{\bibfnamefont{P.}~\bibnamefont{Carninci}},
  \bibinfo{author}{\bibfnamefont{Y.}~\bibnamefont{Nishiyama}},
  \bibinfo{author}{\bibfnamefont{A.}~\bibnamefont{Westover}},
  \bibinfo{author}{\bibfnamefont{M.}~\bibnamefont{Itoh}},
  \bibinfo{author}{\bibfnamefont{S.}~\bibnamefont{Nagaoka}},
  \bibinfo{author}{\bibfnamefont{N.}~\bibnamefont{Sasaki}},
  \bibinfo{author}{\bibfnamefont{Y.}~\bibnamefont{Okazaki}},
  \bibinfo{author}{\bibfnamefont{M.}~\bibnamefont{Muramatsu}},
  \bibnamefont{and}
  \bibinfo{author}{\bibfnamefont{Y.}~\bibnamefont{Hayashizaki}},
  \bibinfo{journal}{Proc. Natl. Acad. Sci. USA} \textbf{\bibinfo{volume}{95}},
  \bibinfo{pages}{520} (\bibinfo{year}{1998}).

\bibitem[{\citenamefont{Sola-Penna and Meyer-Fernandes}(1998)}]{mey98a}
\bibinfo{author}{\bibfnamefont{M.}~\bibnamefont{Sola-Penna}} \bibnamefont{and}
  \bibinfo{author}{\bibfnamefont{J.~R.} \bibnamefont{Meyer-Fernandes}},
  \bibinfo{journal}{Arch. Biochem. Biophys.} \textbf{\bibinfo{volume}{360}},
  \bibinfo{pages}{10} (\bibinfo{year}{1998}).

\bibitem[{\citenamefont{Kaushik and Bhat}(2003)}]{bha03a}
\bibinfo{author}{\bibfnamefont{J.~K.} \bibnamefont{Kaushik}} \bibnamefont{and}
  \bibinfo{author}{\bibfnamefont{R.}~\bibnamefont{Bhat}}, \bibinfo{journal}{J.
  Biol. Chem.} \textbf{\bibinfo{volume}{278}}, \bibinfo{pages}{26458}
  (\bibinfo{year}{2003}).

\bibitem[{\citenamefont{Miller et~al.}(1997)\citenamefont{Miller, {de Pablo},
  and Corti}}]{cor97a}
\bibinfo{author}{\bibfnamefont{D.~P.} \bibnamefont{Miller}},
  \bibinfo{author}{\bibfnamefont{J.~J.} \bibnamefont{{de Pablo}}},
  \bibnamefont{and} \bibinfo{author}{\bibfnamefont{H.}~\bibnamefont{Corti}},
  \bibinfo{journal}{Pharm. Res.} \textbf{\bibinfo{volume}{14}},
  \bibinfo{pages}{578} (\bibinfo{year}{1997}).

\bibitem[{\citenamefont{Heyden et~al.}(2008)\citenamefont{Heyden,
  Br{\"u}ndemann, Heugen, Niehues, Leitner, and Havenith}}]{hav08a}
\bibinfo{author}{\bibfnamefont{M.}~\bibnamefont{Heyden}},
  \bibinfo{author}{\bibfnamefont{E.}~\bibnamefont{Br{\"u}ndemann}},
  \bibinfo{author}{\bibfnamefont{U.}~\bibnamefont{Heugen}},
  \bibinfo{author}{\bibfnamefont{G.}~\bibnamefont{Niehues}},
  \bibinfo{author}{\bibfnamefont{D.~M.} \bibnamefont{Leitner}},
  \bibnamefont{and} \bibinfo{author}{\bibfnamefont{M.}~\bibnamefont{Havenith}},
  \bibinfo{journal}{J. Am. Chem. Soc.} \textbf{\bibinfo{volume}{130}},
  \bibinfo{pages}{5773} (\bibinfo{year}{2008}).

\bibitem[{\citenamefont{Winther et~al.}(2012)\citenamefont{Winther, Qvist, and
  Halle}}]{hal12a}
\bibinfo{author}{\bibfnamefont{L.~R.} \bibnamefont{Winther}},
  \bibinfo{author}{\bibfnamefont{J.}~\bibnamefont{Qvist}}, \bibnamefont{and}
  \bibinfo{author}{\bibfnamefont{B.}~\bibnamefont{Halle}}, \bibinfo{journal}{J.
  Phys. Chem. B} \textbf{\bibinfo{volume}{116}}, \bibinfo{pages}{9196}
  (\bibinfo{year}{2012}).

\bibitem[{\citenamefont{Crowe}(1971)}]{cro71a}
\bibinfo{author}{\bibfnamefont{J.~H.} \bibnamefont{Crowe}},
  \bibinfo{journal}{American Naturalist} \textbf{\bibinfo{volume}{105}},
  \bibinfo{pages}{563} (\bibinfo{year}{1971}).

\bibitem[{\citenamefont{Olsson et~al.}(2016)\citenamefont{Olsson, Jansson, and
  Swenson}}]{swe16a}
\bibinfo{author}{\bibfnamefont{C.}~\bibnamefont{Olsson}},
  \bibinfo{author}{\bibfnamefont{H.}~\bibnamefont{Jansson}}, \bibnamefont{and}
  \bibinfo{author}{\bibfnamefont{J.}~\bibnamefont{Swenson}},
  \bibinfo{journal}{J. Phys. Chem. B} \textbf{\bibinfo{volume}{120}},
  \bibinfo{pages}{4723} (\bibinfo{year}{2016}).

\bibitem[{\citenamefont{Roy et~al.}(2016)\citenamefont{Roy, Dutta, Kundu,
  Banik, and Sarkar}}]{sar16a}
\bibinfo{author}{\bibfnamefont{A.}~\bibnamefont{Roy}},
  \bibinfo{author}{\bibfnamefont{R.}~\bibnamefont{Dutta}},
  \bibinfo{author}{\bibfnamefont{N.}~\bibnamefont{Kundu}},
  \bibinfo{author}{\bibfnamefont{D.}~\bibnamefont{Banik}}, \bibnamefont{and}
  \bibinfo{author}{\bibfnamefont{N.}~\bibnamefont{Sarkar}},
  \bibinfo{journal}{Langmuir} \textbf{\bibinfo{volume}{32}},
  \bibinfo{pages}{5124} (\bibinfo{year}{2016}).

\bibitem[{\citenamefont{Magaz\`{u} et~al.}(1997)\citenamefont{Magaz\`{u},
  Migliardo, Musolino, and Sciortino}}]{sci97a}
\bibinfo{author}{\bibfnamefont{S.}~\bibnamefont{Magaz\`{u}}},
  \bibinfo{author}{\bibfnamefont{P.}~\bibnamefont{Migliardo}},
  \bibinfo{author}{\bibfnamefont{A.~M.} \bibnamefont{Musolino}},
  \bibnamefont{and} \bibinfo{author}{\bibfnamefont{M.~T.}
  \bibnamefont{Sciortino}}, \bibinfo{journal}{J. Phys. Chem. B}
  \textbf{\bibinfo{volume}{101}}, \bibinfo{pages}{2348} (\bibinfo{year}{1997}).

\bibitem[{\citenamefont{Corradini et~al.}(2013)\citenamefont{Corradini,
  Strekalova, Stanley, and Gallo}}]{gal13a}
\bibinfo{author}{\bibfnamefont{D.}~\bibnamefont{Corradini}},
  \bibinfo{author}{\bibfnamefont{E.~G.} \bibnamefont{Strekalova}},
  \bibinfo{author}{\bibfnamefont{H.~E.} \bibnamefont{Stanley}},
  \bibnamefont{and} \bibinfo{author}{\bibfnamefont{P.}~\bibnamefont{Gallo}},
  \bibinfo{journal}{Scientific Reports} \textbf{\bibinfo{volume}{3}},
  \bibinfo{pages}{1218} (\bibinfo{year}{2013}).

\bibitem[{\citenamefont{Shukla et~al.}(2016)\citenamefont{Shukla, Pomarico,
  Chen, Chergui, and Othon}}]{oth16a}
\bibinfo{author}{\bibfnamefont{N.}~\bibnamefont{Shukla}},
  \bibinfo{author}{\bibfnamefont{E.}~\bibnamefont{Pomarico}},
  \bibinfo{author}{\bibfnamefont{L.}~\bibnamefont{Chen}},
  \bibinfo{author}{\bibfnamefont{M.}~\bibnamefont{Chergui}}, \bibnamefont{and}
  \bibinfo{author}{\bibfnamefont{C.~M.} \bibnamefont{Othon}},
  \bibinfo{journal}{J. Phys. Chem. B} \textbf{\bibinfo{volume}{120}},
  \bibinfo{pages}{9477} (\bibinfo{year}{2016}).

\bibitem[{\citenamefont{Sajadi et~al.}(2014)\citenamefont{Sajadi, Berndt,
  Richter, Gerecke, Mahrwald, and Ernsting}}]{ern14a}
\bibinfo{author}{\bibfnamefont{M.}~\bibnamefont{Sajadi}},
  \bibinfo{author}{\bibfnamefont{F.}~\bibnamefont{Berndt}},
  \bibinfo{author}{\bibfnamefont{C.}~\bibnamefont{Richter}},
  \bibinfo{author}{\bibfnamefont{M.}~\bibnamefont{Gerecke}},
  \bibinfo{author}{\bibfnamefont{R.}~\bibnamefont{Mahrwald}}, \bibnamefont{and}
  \bibinfo{author}{\bibfnamefont{N.~P.} \bibnamefont{Ernsting}},
  \bibinfo{journal}{J. Phys. Chem. Lett.} \textbf{\bibinfo{volume}{5}},
  \bibinfo{pages}{1845} (\bibinfo{year}{2014}).

\bibitem[{\citenamefont{Maroncelli and Fleming}(1987)}]{fle87a}
\bibinfo{author}{\bibfnamefont{M.}~\bibnamefont{Maroncelli}} \bibnamefont{and}
  \bibinfo{author}{\bibfnamefont{G.~R.} \bibnamefont{Fleming}},
  \bibinfo{journal}{J. Chem. Phys.} \textbf{\bibinfo{volume}{86}},
  \bibinfo{pages}{6221} (\bibinfo{year}{1987}).

\bibitem[{\citenamefont{Vajda et~al.}(1995)\citenamefont{Vajda, Jimenez,
  Rosenthal, Fidler, Fleming, and {Castner, Jr.}}}]{cas95a}
\bibinfo{author}{\bibfnamefont{S.}~\bibnamefont{Vajda}},
  \bibinfo{author}{\bibfnamefont{R.}~\bibnamefont{Jimenez}},
  \bibinfo{author}{\bibfnamefont{S.~J.} \bibnamefont{Rosenthal}},
  \bibinfo{author}{\bibfnamefont{V.}~\bibnamefont{Fidler}},
  \bibinfo{author}{\bibfnamefont{G.~R.} \bibnamefont{Fleming}},
  \bibnamefont{and} \bibinfo{author}{\bibfnamefont{E.~W.}
  \bibnamefont{{Castner, Jr.}}}, \bibinfo{journal}{J. Chem. Soc. Faraday
  Trans.} \textbf{\bibinfo{volume}{91}}, \bibinfo{pages}{867}
  (\bibinfo{year}{1995}).

\bibitem[{\citenamefont{Maroncelli}(1993)}]{mar93a}
\bibinfo{author}{\bibfnamefont{M.}~\bibnamefont{Maroncelli}},
  \bibinfo{journal}{J. Mol. Liquids} \textbf{\bibinfo{volume}{57}},
  \bibinfo{pages}{1} (\bibinfo{year}{1993}).

\bibitem[{\citenamefont{Mertz et~al.}(1997)\citenamefont{Mertz, Tikhomirov, and
  Krishtalik}}]{kri97a}
\bibinfo{author}{\bibfnamefont{E.~L.} \bibnamefont{Mertz}},
  \bibinfo{author}{\bibfnamefont{V.~A.} \bibnamefont{Tikhomirov}},
  \bibnamefont{and} \bibinfo{author}{\bibfnamefont{L.~I.}
  \bibnamefont{Krishtalik}}, \bibinfo{journal}{J. Phys. Chem.}
  \textbf{\bibinfo{volume}{101}}, \bibinfo{pages}{3433} (\bibinfo{year}{1997}).

\bibitem[{\citenamefont{Nilsson and Halle}(2005)}]{hal05a}
\bibinfo{author}{\bibfnamefont{L.}~\bibnamefont{Nilsson}} \bibnamefont{and}
  \bibinfo{author}{\bibfnamefont{B.}~\bibnamefont{Halle}},
  \bibinfo{journal}{PNAS} \textbf{\bibinfo{volume}{102}},
  \bibinfo{pages}{13967} (\bibinfo{year}{2005}).

\bibitem[{\citenamefont{Guang-Yu et~al.}(2015)\citenamefont{Guang-Yu, Yu, Wei,
  Shu-Feng, Zhong, and Qi-Huang}}]{qih15a}
\bibinfo{author}{\bibfnamefont{G.}~\bibnamefont{Guang-Yu}},
  \bibinfo{author}{\bibfnamefont{L.}~\bibnamefont{Yu}},
  \bibinfo{author}{\bibfnamefont{W.}~\bibnamefont{Wei}},
  \bibinfo{author}{\bibfnamefont{W.}~\bibnamefont{Shu-Feng}},
  \bibinfo{author}{\bibfnamefont{D.}~\bibnamefont{Zhong}}, \bibnamefont{and}
  \bibinfo{author}{\bibfnamefont{G.}~\bibnamefont{Qi-Huang}},
  \bibinfo{journal}{Chin. Phys. B} \textbf{\bibinfo{volume}{24}},
  \bibinfo{pages}{018201} (\bibinfo{year}{2015}).

\bibitem[{\citenamefont{Pal et~al.}(2015)\citenamefont{Pal, Shweta, Singh,
  Verma, and Sen}}]{sen15a}
\bibinfo{author}{\bibfnamefont{N.}~\bibnamefont{Pal}},
  \bibinfo{author}{\bibfnamefont{H.}~\bibnamefont{Shweta}},
  \bibinfo{author}{\bibfnamefont{M.~K.} \bibnamefont{Singh}},
  \bibinfo{author}{\bibfnamefont{S.~D.} \bibnamefont{Verma}}, \bibnamefont{and}
  \bibinfo{author}{\bibfnamefont{S.}~\bibnamefont{Sen}}, \bibinfo{journal}{J.
  Phys. Chem. Lett.} \textbf{\bibinfo{volume}{6}}, \bibinfo{pages}{1754}
  (\bibinfo{year}{2015}).

\bibitem[{\citenamefont{Yadav et~al.}(2016)\citenamefont{Yadav, Sengupta, and
  Sen}}]{sen16a}
\bibinfo{author}{\bibfnamefont{R.}~\bibnamefont{Yadav}},
  \bibinfo{author}{\bibfnamefont{B.}~\bibnamefont{Sengupta}}, \bibnamefont{and}
  \bibinfo{author}{\bibfnamefont{P.}~\bibnamefont{Sen}},
  \bibinfo{journal}{Biophys. Chem.} \textbf{\bibinfo{volume}{211}},
  \bibinfo{pages}{59} (\bibinfo{year}{2016}).

\bibitem[{\citenamefont{Sen et~al.}(2009)\citenamefont{Sen, Andreatta,
  Ponomarev, Beveridge, and Berg}}]{ber09a}
\bibinfo{author}{\bibfnamefont{S.}~\bibnamefont{Sen}},
  \bibinfo{author}{\bibfnamefont{D.}~\bibnamefont{Andreatta}},
  \bibinfo{author}{\bibfnamefont{S.~Y.} \bibnamefont{Ponomarev}},
  \bibinfo{author}{\bibfnamefont{D.~L.} \bibnamefont{Beveridge}},
  \bibnamefont{and} \bibinfo{author}{\bibfnamefont{M.~A.} \bibnamefont{Berg}},
  \bibinfo{journal}{J. Am. Chem. Soc.} \textbf{\bibinfo{volume}{131}},
  \bibinfo{pages}{1724} (\bibinfo{year}{2009}).

\bibitem[{\citenamefont{Sajadi et~al.}(2010)\citenamefont{Sajadi, Ajaj, Ioffe,
  Weing{\"a}rtner, and Ernsting}}]{ern10a}
\bibinfo{author}{\bibfnamefont{M.}~\bibnamefont{Sajadi}},
  \bibinfo{author}{\bibfnamefont{Y.}~\bibnamefont{Ajaj}},
  \bibinfo{author}{\bibfnamefont{I.}~\bibnamefont{Ioffe}},
  \bibinfo{author}{\bibfnamefont{H.}~\bibnamefont{Weing{\"a}rtner}},
  \bibnamefont{and} \bibinfo{author}{\bibfnamefont{N.~P.}
  \bibnamefont{Ernsting}}, \bibinfo{journal}{Angew. Chem. Int. Ed.}
  \textbf{\bibinfo{volume}{49}}, \bibinfo{pages}{454} (\bibinfo{year}{2010}).

\bibitem[{\citenamefont{Verde and Campen}(2011)}]{cam11a}
\bibinfo{author}{\bibfnamefont{A.~V.} \bibnamefont{Verde}} \bibnamefont{and}
  \bibinfo{author}{\bibfnamefont{R.~K.} \bibnamefont{Campen}},
  \bibinfo{journal}{J. Phys. Chem. B} \textbf{\bibinfo{volume}{115}},
  \bibinfo{pages}{7069} (\bibinfo{year}{2011}).

\bibitem[{\citenamefont{Lamoureux and Roux}(2003)}]{rou03a}
\bibinfo{author}{\bibfnamefont{G.}~\bibnamefont{Lamoureux}} \bibnamefont{and}
  \bibinfo{author}{\bibfnamefont{B.}~\bibnamefont{Roux}}, \bibinfo{journal}{J.
  Chem. Phys.} \textbf{\bibinfo{volume}{119}}, \bibinfo{pages}{3025}
  (\bibinfo{year}{2003}).

\bibitem[{\citenamefont{Heid et~al.}(2016)\citenamefont{Heid, Harringer, and
  Schr\"oder}}]{sch16b}
\bibinfo{author}{\bibfnamefont{E.}~\bibnamefont{Heid}},
  \bibinfo{author}{\bibfnamefont{S.}~\bibnamefont{Harringer}},
  \bibnamefont{and}
  \bibinfo{author}{\bibfnamefont{C.}~\bibnamefont{Schr\"oder}},
  \bibinfo{journal}{J. Chem. Phys.} \bibinfo{pages}{accepted}
  (\bibinfo{year}{2016}).

\bibitem[{\citenamefont{Chai and Head-Gordon}(2008)}]{cha08a}
\bibinfo{author}{\bibfnamefont{J.-D.} \bibnamefont{Chai}} \bibnamefont{and}
  \bibinfo{author}{\bibfnamefont{M.}~\bibnamefont{Head-Gordon}},
  \bibinfo{journal}{Phys. Chem. Chem. Phys.} \textbf{\bibinfo{volume}{10}},
  \bibinfo{pages}{6615} (\bibinfo{year}{2008}).

\bibitem[{\citenamefont{Tomasi et~al.}(2005)\citenamefont{Tomasi, Mennucci, and
  Cammi}}]{tom05b}
\bibinfo{author}{\bibfnamefont{J.}~\bibnamefont{Tomasi}},
  \bibinfo{author}{\bibfnamefont{B.}~\bibnamefont{Mennucci}}, \bibnamefont{and}
  \bibinfo{author}{\bibfnamefont{R.}~\bibnamefont{Cammi}},
  \bibinfo{journal}{Chem. Rev} \textbf{\bibinfo{volume}{105}},
  \bibinfo{pages}{2999} (\bibinfo{year}{2005}).

\bibitem[{\citenamefont{Vanommeslaeghe and {MacKerell Jr.}}(2012)}]{van12a}
\bibinfo{author}{\bibfnamefont{K.}~\bibnamefont{Vanommeslaeghe}}
  \bibnamefont{and} \bibinfo{author}{\bibfnamefont{A.~D.}
  \bibnamefont{{MacKerell Jr.}}}, \bibinfo{journal}{J. Chem. Inf. Model.}
  \textbf{\bibinfo{volume}{52}}, \bibinfo{pages}{3144} (\bibinfo{year}{2012}).

\bibitem[{\citenamefont{Vanommeslaeghe
  et~al.}(2012)\citenamefont{Vanommeslaeghe, Raman, and {MacKerell
  Jr.}}}]{van12b}
\bibinfo{author}{\bibfnamefont{K.}~\bibnamefont{Vanommeslaeghe}},
  \bibinfo{author}{\bibfnamefont{E.~P.} \bibnamefont{Raman}}, \bibnamefont{and}
  \bibinfo{author}{\bibfnamefont{A.~D.} \bibnamefont{{MacKerell Jr.}}},
  \bibinfo{journal}{J. Chem. Inf. Model.} \textbf{\bibinfo{volume}{52}},
  \bibinfo{pages}{3155} (\bibinfo{year}{2012}).

\bibitem[{\citenamefont{Vanommeslaeghe
  et~al.}(2010)\citenamefont{Vanommeslaeghe, Hatcher, Acharya, Kundu, Zhong,
  Shim, Darian, Guvench, Lopes, Vorobyov et~al.}}]{van10a}
\bibinfo{author}{\bibfnamefont{K.}~\bibnamefont{Vanommeslaeghe}},
  \bibinfo{author}{\bibfnamefont{E.}~\bibnamefont{Hatcher}},
  \bibinfo{author}{\bibfnamefont{C.}~\bibnamefont{Acharya}},
  \bibinfo{author}{\bibfnamefont{S.}~\bibnamefont{Kundu}},
  \bibinfo{author}{\bibfnamefont{S.}~\bibnamefont{Zhong}},
  \bibinfo{author}{\bibfnamefont{J.}~\bibnamefont{Shim}},
  \bibinfo{author}{\bibfnamefont{E.}~\bibnamefont{Darian}},
  \bibinfo{author}{\bibfnamefont{O.}~\bibnamefont{Guvench}},
  \bibinfo{author}{\bibfnamefont{P.}~\bibnamefont{Lopes}},
  \bibinfo{author}{\bibfnamefont{I.}~\bibnamefont{Vorobyov}},
  \bibnamefont{et~al.}, \bibinfo{journal}{J. Comp. Chem.}
  \textbf{\bibinfo{volume}{31}}, \bibinfo{pages}{671} (\bibinfo{year}{2010}).

\bibitem[{\citenamefont{Brooks et~al.}(2009)\citenamefont{Brooks, {Brooks III},
  {MacKerell Jr.}, Nilsson, Petrella, Roux, Won, Archontis, Bartels, Boresch
  et~al.}}]{kar09a}
\bibinfo{author}{\bibfnamefont{B.~R.} \bibnamefont{Brooks}},
  \bibinfo{author}{\bibfnamefont{C.~L.} \bibnamefont{{Brooks III}}},
  \bibinfo{author}{\bibfnamefont{A.~D.} \bibnamefont{{MacKerell Jr.}}},
  \bibinfo{author}{\bibfnamefont{L.}~\bibnamefont{Nilsson}},
  \bibinfo{author}{\bibfnamefont{R.~J.} \bibnamefont{Petrella}},
  \bibinfo{author}{\bibfnamefont{B.}~\bibnamefont{Roux}},
  \bibinfo{author}{\bibfnamefont{Y.}~\bibnamefont{Won}},
  \bibinfo{author}{\bibfnamefont{G.}~\bibnamefont{Archontis}},
  \bibinfo{author}{\bibfnamefont{C.}~\bibnamefont{Bartels}},
  \bibinfo{author}{\bibfnamefont{S.}~\bibnamefont{Boresch}},
  \bibnamefont{et~al.}, \bibinfo{journal}{J. Comput. Chem.}
  \textbf{\bibinfo{volume}{30}}, \bibinfo{pages}{1545} (\bibinfo{year}{2009}).

\bibitem[{\citenamefont{Neumayr et~al.}(2009)\citenamefont{Neumayr,
  Schr{\"o}der, and Steinhauser}}]{ste09d}
\bibinfo{author}{\bibfnamefont{G.}~\bibnamefont{Neumayr}},
  \bibinfo{author}{\bibfnamefont{C.}~\bibnamefont{Schr{\"o}der}},
  \bibnamefont{and}
  \bibinfo{author}{\bibfnamefont{O.}~\bibnamefont{Steinhauser}},
  \bibinfo{journal}{J. Chem. Phys.} \textbf{\bibinfo{volume}{131}},
  \bibinfo{pages}{174509} (\bibinfo{year}{2009}).

\bibitem[{\citenamefont{Okabe}(2000)}]{oka00a}
\bibinfo{author}{\bibfnamefont{A.}~\bibnamefont{Okabe}},
  \emph{\bibinfo{title}{Spatial tesselations: {C}oncepts and applications of
  {V}oronoi diagrams}} (\bibinfo{publisher}{Wiley}, \bibinfo{address}{New
  York}, \bibinfo{year}{2000}).

\bibitem[{\citenamefont{Michaud-Agrawal
  et~al.}(2011)\citenamefont{Michaud-Agrawal, Denning, Woolf, and
  Beckstein}}]{bec11a}
\bibinfo{author}{\bibfnamefont{N.}~\bibnamefont{Michaud-Agrawal}},
  \bibinfo{author}{\bibfnamefont{E.~J.} \bibnamefont{Denning}},
  \bibinfo{author}{\bibfnamefont{T.~B.} \bibnamefont{Woolf}}, \bibnamefont{and}
  \bibinfo{author}{\bibfnamefont{O.}~\bibnamefont{Beckstein}},
  \bibinfo{journal}{J. Comput. Chem.} \textbf{\bibinfo{volume}{32}},
  \bibinfo{pages}{2319} (\bibinfo{year}{2011}).

\bibitem[{\citenamefont{Lerbet et~al.}(2011)\citenamefont{Lerbet, Affouard,
  Bordat, H\'edoux, Guinet, and Descamps}}]{des11a}
\bibinfo{author}{\bibfnamefont{A.}~\bibnamefont{Lerbet}},
  \bibinfo{author}{\bibfnamefont{F.}~\bibnamefont{Affouard}},
  \bibinfo{author}{\bibfnamefont{P.}~\bibnamefont{Bordat}},
  \bibinfo{author}{\bibfnamefont{A.}~\bibnamefont{H\'edoux}},
  \bibinfo{author}{\bibfnamefont{Y.}~\bibnamefont{Guinet}}, \bibnamefont{and}
  \bibinfo{author}{\bibfnamefont{M.}~\bibnamefont{Descamps}},
  \bibinfo{journal}{J. Non.-Cryst. Sol.} \textbf{\bibinfo{volume}{357}},
  \bibinfo{pages}{695} (\bibinfo{year}{2011}).

\bibitem[{\citenamefont{Lupi et~al.}(2012)\citenamefont{Lupi, Comez,
  Paolantoni, Perticaroli, Sassi, Morresi, Ladanyi, and Fioretto}}]{fio12a}
\bibinfo{author}{\bibfnamefont{L.}~\bibnamefont{Lupi}},
  \bibinfo{author}{\bibfnamefont{L.}~\bibnamefont{Comez}},
  \bibinfo{author}{\bibfnamefont{M.}~\bibnamefont{Paolantoni}},
  \bibinfo{author}{\bibfnamefont{S.}~\bibnamefont{Perticaroli}},
  \bibinfo{author}{\bibfnamefont{P.}~\bibnamefont{Sassi}},
  \bibinfo{author}{\bibfnamefont{A.}~\bibnamefont{Morresi}},
  \bibinfo{author}{\bibfnamefont{B.~M.} \bibnamefont{Ladanyi}},
  \bibnamefont{and} \bibinfo{author}{\bibfnamefont{D.}~\bibnamefont{Fioretto}},
  \bibinfo{journal}{J. Phys. Chem. B} \textbf{\bibinfo{volume}{116}},
  \bibinfo{pages}{14760} (\bibinfo{year}{2012}).

\bibitem[{\citenamefont{Holz et~al.}(2000)\citenamefont{Holz, Heil, and
  Sacco}}]{sac00a}
\bibinfo{author}{\bibfnamefont{M.}~\bibnamefont{Holz}},
  \bibinfo{author}{\bibfnamefont{S.~R.} \bibnamefont{Heil}}, \bibnamefont{and}
  \bibinfo{author}{\bibfnamefont{A.}~\bibnamefont{Sacco}},
  \bibinfo{journal}{Phys. Chem. Chem. Phys.} \textbf{\bibinfo{volume}{2}},
  \bibinfo{pages}{4740} (\bibinfo{year}{2000}).

\bibitem[{\citenamefont{Bagchi and Jana}(2010)}]{jan10a}
\bibinfo{author}{\bibfnamefont{B.}~\bibnamefont{Bagchi}} \bibnamefont{and}
  \bibinfo{author}{\bibfnamefont{B.}~\bibnamefont{Jana}},
  \bibinfo{journal}{Chem. Soc. Rev.} \textbf{\bibinfo{volume}{39}},
  \bibinfo{pages}{1936} (\bibinfo{year}{2010}).

\bibitem[{\citenamefont{Jimenez et~al.}(1994)\citenamefont{Jimenez, Fleming,
  Kumar, and Maroncelli}}]{mar94a}
\bibinfo{author}{\bibfnamefont{R.}~\bibnamefont{Jimenez}},
  \bibinfo{author}{\bibfnamefont{G.~R.} \bibnamefont{Fleming}},
  \bibinfo{author}{\bibfnamefont{P.~V.} \bibnamefont{Kumar}}, \bibnamefont{and}
  \bibinfo{author}{\bibfnamefont{M.}~\bibnamefont{Maroncelli}},
  \bibinfo{journal}{Nature} \textbf{\bibinfo{volume}{369}},
  \bibinfo{pages}{471} (\bibinfo{year}{1994}).

\bibitem[{\citenamefont{Roy and Maroncelli}(2012)}]{mar12b}
\bibinfo{author}{\bibfnamefont{D.}~\bibnamefont{Roy}} \bibnamefont{and}
  \bibinfo{author}{\bibfnamefont{M.}~\bibnamefont{Maroncelli}},
  \bibinfo{journal}{J. Phys. Chem. B} \textbf{\bibinfo{volume}{116}},
  \bibinfo{pages}{5951} (\bibinfo{year}{2012}).

\bibitem[{\citenamefont{Chowdhury et~al.}(2004)\citenamefont{Chowdhury, Halder,
  Sanders, Calhoun, Anderson, Armstrong, Song, and Petrich}}]{pet04a}
\bibinfo{author}{\bibfnamefont{P.~K.} \bibnamefont{Chowdhury}},
  \bibinfo{author}{\bibfnamefont{M.}~\bibnamefont{Halder}},
  \bibinfo{author}{\bibfnamefont{L.}~\bibnamefont{Sanders}},
  \bibinfo{author}{\bibfnamefont{T.}~\bibnamefont{Calhoun}},
  \bibinfo{author}{\bibfnamefont{J.~L.} \bibnamefont{Anderson}},
  \bibinfo{author}{\bibfnamefont{D.~W.} \bibnamefont{Armstrong}},
  \bibinfo{author}{\bibfnamefont{X.}~\bibnamefont{Song}}, \bibnamefont{and}
  \bibinfo{author}{\bibfnamefont{J.~W.} \bibnamefont{Petrich}},
  \bibinfo{journal}{J. Phys. Chem. B} \textbf{\bibinfo{volume}{108}},
  \bibinfo{pages}{10245} (\bibinfo{year}{2004}).

\bibitem[{\citenamefont{Karmakar and Samanta}(2002)}]{sam02a}
\bibinfo{author}{\bibfnamefont{R.}~\bibnamefont{Karmakar}} \bibnamefont{and}
  \bibinfo{author}{\bibfnamefont{A.}~\bibnamefont{Samanta}},
  \bibinfo{journal}{J. Phys. Chem. A} \textbf{\bibinfo{volume}{106}},
  \bibinfo{pages}{4447} (\bibinfo{year}{2002}).

\bibitem[{\citenamefont{Karmakar and Samanta}(2003)}]{sam03a}
\bibinfo{author}{\bibfnamefont{R.}~\bibnamefont{Karmakar}} \bibnamefont{and}
  \bibinfo{author}{\bibfnamefont{A.}~\bibnamefont{Samanta}},
  \bibinfo{journal}{J. Phys. Chem. A} \textbf{\bibinfo{volume}{107}},
  \bibinfo{pages}{7340} (\bibinfo{year}{2003}).

\bibitem[{\citenamefont{Groot and Baker}(2015)}]{bak15a}
\bibinfo{author}{\bibfnamefont{C.~C.~M.} \bibnamefont{Groot}} \bibnamefont{and}
  \bibinfo{author}{\bibfnamefont{H.~J.} \bibnamefont{Baker}},
  \bibinfo{journal}{Phys. Chem. Chem. Phys} \textbf{\bibinfo{volume}{17}},
  \bibinfo{pages}{9440} (\bibinfo{year}{2015}).

\bibitem[{\citenamefont{Shiraga et~al.}(2015)\citenamefont{Shiraga, Suzuki,
  Kondo, Baerdemaeker, and Ogawa}}]{oga15a}
\bibinfo{author}{\bibfnamefont{K.}~\bibnamefont{Shiraga}},
  \bibinfo{author}{\bibfnamefont{T.}~\bibnamefont{Suzuki}},
  \bibinfo{author}{\bibfnamefont{N.}~\bibnamefont{Kondo}},
  \bibinfo{author}{\bibfnamefont{J.~D.} \bibnamefont{Baerdemaeker}},
  \bibnamefont{and} \bibinfo{author}{\bibfnamefont{Y.}~\bibnamefont{Ogawa}},
  \bibinfo{journal}{Carbohydrate Research} \textbf{\bibinfo{volume}{406}},
  \bibinfo{pages}{46} (\bibinfo{year}{2015}).

\bibitem[{\citenamefont{Pagnotta et~al.}(2010)\citenamefont{Pagnotta, McLain,
  Soper, Bruni, and Ricci}}]{ric10a}
\bibinfo{author}{\bibfnamefont{S.~E.} \bibnamefont{Pagnotta}},
  \bibinfo{author}{\bibfnamefont{S.~E.} \bibnamefont{McLain}},
  \bibinfo{author}{\bibfnamefont{A.~K.} \bibnamefont{Soper}},
  \bibinfo{author}{\bibfnamefont{F.}~\bibnamefont{Bruni}}, \bibnamefont{and}
  \bibinfo{author}{\bibfnamefont{M.~A.} \bibnamefont{Ricci}},
  \bibinfo{journal}{J. Phys. Chem} \textbf{\bibinfo{volume}{114}},
  \bibinfo{pages}{4904} (\bibinfo{year}{2010}).

\end{thebibliography}
\end{document}